Merging or division. I. Dependencies between the magnitude gap of groups and the morphology of two brightest galaxies

A.P. Mahtessian, V.H. Movsisyan, L.A. Mahtessian
Byurakan Astrophysical Observatory, amahtes@bao.sci.am

*1. Introduction.* The so-called "fossil groups of galaxies" contain of a giant elliptical galaxy loaded in an X-ray emitting gaseous medium commensurate with group size. It is believed that they are the result of a multiple mergers of group members. The luminosity of central elliptical galaxy of the fossil group is characteristic central galaxy of clusters, but does not have a vast stellar halo characteristic of these galaxies. Despite the existence of these giant galaxies, the luminous matter of the fossil group is only 10% of the mass of the system. It is supposed that the rest of the mass, in the form of a dark matter, is necessary to maintain X-ray radiation of gas gravity-associated with the system.

The study of fossil groups began in the recent past. Such as, the first group is represented by Ponman et al. (1994). They defined fossil groups as systems that have a bright ($> 5 \times 10^{43} h_{50}^{-2} erg\, s^{-1}$) and extended X-ray halo, the dominant elliptical galaxy with the environment of weak satellite galaxies. The difference between the magnitudes of the first and second ranked galaxies, m12, should exceed two.

Galactic systems with a large spread of m12 are observed in the nature. Since it is considered that the difference m12 is associated with the mergers of galaxies, this parameter should increase with each merger. That is, it characterizes the dynamic age of the system. Therefore, fossil and non-fossil groups can have different galactic populations. This means that they will also differ in their luminosity functions. LF was studied in Zarattini et al. (2015). In this paper, a luminosity function is constructed for 102 systems with redshifts z≤0.25. It was found that the systems with large differences m12 have small values of M*, i.e., in these groups a small relative number of bright galaxies are observed. Groups with different values of m12 also differed by a weak end of luminosity function. For larger gap, a flatter luminosity function is observed, i.e. in these groups a small relative number of weak galaxies are also observed. The behavior of the LF in the bright end is consistent with the hypothesis of merger, and the behavior of the weak end - is not. The authors of the article assume that observational errors can play a role here.

So far, works relating to fossil groups are inclined to prove that in these groups the main mechanism of evolution of galaxies is associated with the merging of galaxies.

It should be noted that there is another hypothesis proposed by V. Ambartsumyan in the Byurakan Observatory, in which the evolution of galaxies is interpreted in a diametrically opposite way (V.A. Ambartsumian 1956, 1964).

Many observational facts can be explained by both mechanisms, but the dependencies of the characteristics of galaxies from the magnitude gap of the groups might enable us to understand which of these two opposite mechanisms, in fact, makes a big contribution to the evolution of galaxies.

As the first characteristic, we can take the observed morphological type of galaxies. According to the merging scenario, the mass (or luminosity) of the central galaxy of the group must grow with time, and the central galaxy must take an elliptical form. According to Ambartsumian's version of the evolution of galaxies, no assumption is made about the morphology of the central galaxy. According to Ambartsumian, there are two possible variants of the origin of galaxies.

a. At first, the protogroup disintegrates to parts, hereafter from which are formed members of the group.

b. At first, the first-raked galaxy of the group is formed, from which later on by means of ejection (or disintegration) are formed other members of the group.

Both versions lead to the assumption that in groups of galaxies there must be a consistency between the morphological types (See, Mahtessian , Movsessian, 2001).

**Thus, from the point of view of the adherents of the merging, a large magnitude gap of group means a long dynamical evolution. From the point of view of the formation of galaxies from protogroup matter through disintegration, this gap is either not related to the dynamical evolution of the groups or there must be an inverse relationship, i.e., a large gap m12 means a short dynamic age.**

In this paper we consider the dependence between magnitude gap and the morphological types of the first and second ranked galaxies of groups.

*2. Sample.* We have used the list of groups of galaxies suggested by Mahtessian, Movsessian (2010). The study area is limited to the following:

1000 km/s ≤ V ≤ 15000 km/s,
$|bII| \leq 20^o$

*3. Results.* Figure 1 shows the distributions of the morphological types of the first and second ranked group galaxies for different magnitude gap m12. Discussed only those groups in which the morphological types of both most luminous galaxies are known. Morphological types are coded as follows, adopted in the literature:
-7 (Unclassified Elliptical); -6 (Compact Elliptical); -5 (E, and dwarf E); -4 (E/SO); -3 (L-, SO-); -2 (L, SO); -1 (L+, SO+); 0 (SO/a, SO-a); 1 (Sa); 2 (Sab); 3 (Sb); 4 (Sbc); 5 (Sc); 6 (Scd); 7 (Sd); 8 (Sdm); 9 (Sm, Magellanic Spiral); 10 (Im, Irr I, Magellanic Irregular, Dwarf Irregular); 15 (Peculiar, Unclassifiable); 20 (S..., Sc-Irr, Unclassified Spiral).

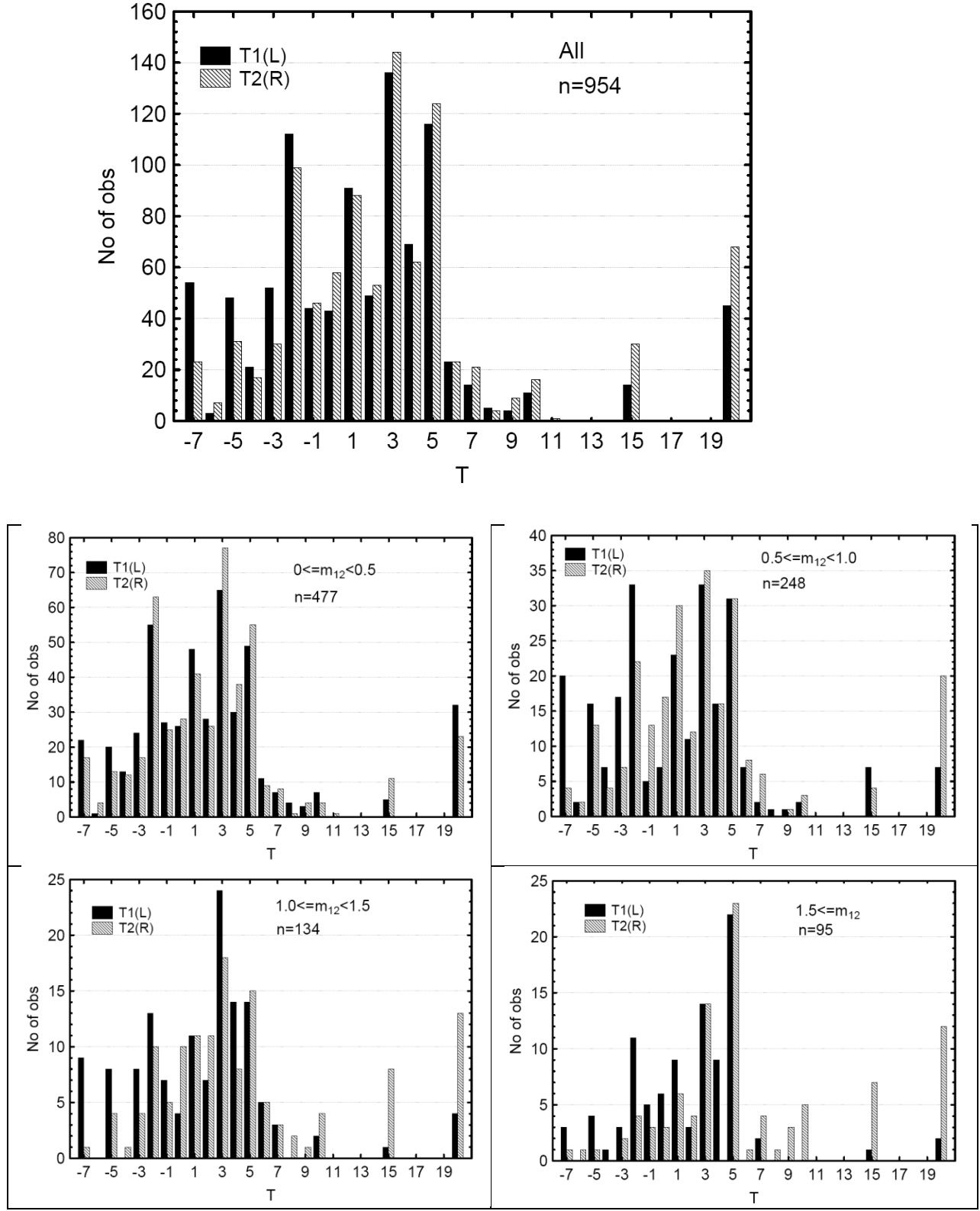

Fig.1. The distribution of the morphological types of the first and second raked group galaxies for different magnitude gap m12.

It is seen from Fig. 1 that the first ranked group galaxies are inclined, more often, to have an elliptical or lenticular shape than the second ranked galaxies of the group, but there is no tendency that this characteristic is more pronounced at large gap m12.

Let us estimate the dependence of the morphological type of a first ranked galaxy on the magnitude gap m12. The results are shown in Table 1. In the table, the galaxies are grouped according to the morphological types of the first ranked galaxies, as follows:

E,L (T=-7÷-1) – elliptical and lenticular galaxies,
S,I (T=0÷20) – Spiral and irregular galaxies.

In the table for a given subsample of galaxies, the observed number, the expected number (in parentheses, based on the fact that there is no correlation between the morphological type of first ranked galaxy and the magnitude gap) are given, and the frequency of appearance m12 is given below. In the last row, regardless of the morphological type, the distributions of the number and the relative number of gap m12 are given.

In the case of elliptical and lenticular galaxies, the statistical significance of the difference between the distributions of the observed and expected numbers is 0.15 by the $\chi^2$ method. In the case of spiral and irregular galaxies, this is ~ 0.4. This means that between these distributions there is no significant statistical difference.

Thus, we can state that the relative number of brightest elliptical galaxy in groups does not increase with increasing magnitude gap m12. This casts doubt on the mechanism of merging.

Table 1. Distributions of magnitude gap in groups, depending on the first ranked galaxy morphology type.

| m12 | 0÷0.5 | 0.5÷1.0 | 1.0÷1.5 | ≥1.5 | All |
|---|---|---|---|---|---|
| E,L (T=-7÷-1) | 217(230.9) 0.45 | 146(125.7) 0.31 | 71(69.1) 0.15 | 41(49.3) 0.09 | 475 |
| S,I (T=0÷20) | 472(458.0) 0.50 | 229(249.3) 0.24 | 135(136.9) 0.14 | 106(97.7) 0.11 | 942 |
| Σ | 689 0.49 | 375 0.26 | 206 0.15 | 147 0.10 | 1417 |

| m12 | ≥2.0 | ≥2.5 | ≥3.0 |
|---|---|---|---|
| E,L (T=-7÷-1) | 17(21.5) 0.036 | 10(10.7) 0.021 | 4(5.0) 0.008 |
| S,I (T=0÷20) | 47(42.5) 0.050 | 22(21.3) 0.023 | 11(10.0) 0.012 |
| Σ | 64 0.045 | 32 0.023 | 15 0.011 |

Table 2. Distributions of magnitude gap in groups, depending on the second ranked galaxy morphology type.

| m12 | 0÷0.5 | 0.5÷1.0 | 1.0÷1.5 | ≥1.5 | All |
|---|---|---|---|---|---|
| E,L (T=-7÷-1) | 173(149.9) 0.63 | 66(67.7) 0.24 | 25(34.1) 0.09 | 12(24.3) 0.04 | 276 |
| S,I (T=0÷20) | 420(443.1) 0.51 | 202(200.3) 0.25 | 110(100.9) 0.13 | 84(71.7) 0.10 | 816 |
| ∑ | 593 0.54 | 268 0.25 | 135 0.12 | 96 0.09 | 1092 |

| m12 | ≥2.0 | ≥2.5 | ≥3.0 |
|---|---|---|---|
| E,L (T=-7÷-1) | 7(11.1) 0.025 | 2(5.6) 0.007 | 1(2.5) 0.004 |
| S,I (T=0÷20) | 37(32.9) 0.045 | 20(16.4) 0.025 | 9(7.5) 0.011 |
| ∑ | 44 0.040 | 22 0.020 | 10 0.009 |

Table 2 shows the distribution of the magnitude gap m12 in groups, depending on the morphological type of the second ranked galaxy. It can be seen from the table that in the transition from groups with small gap to groups with large gap, the number of elliptical and lenticular galaxies decreases with respect to the expected number. The statistical significance of the above, based on the χ2 criterion, is higher (<0.01). This fact cannot be explained by the mechanism of merging.

Let's consider the same question by dividing the sample by m12 into two parts. Table 3 gives the corresponding data for the brightest galaxy for m12 <2.0 and m12≥2.0.

Table 3. The same as table 1 for m12 <2.0 and m12≥2.0

| m12 | 0÷2.0 | ≥2.0 | All |
|---|---|---|---|
| E,L (T=-7÷-1) | 458(453.5) 0.964 | 17(21.5) 0.036 | 475 |
| S,I (T=0÷20) | 895(899.5) 0.950 | 47(42.5) 0.050 | 942 |
| All | 1353 0.955 | 64 0.045 | 1417 |

Table 3 shows that the relative number of elliptical and lenticular galaxies among a first ranked galaxy groups with m12≥2.0 is even smaller than the expected number (but its statistical significance smallish, α≈0.3, estimated by a method χ2, as well as with using a normal approximation).

Table 4. The same as table 2 for m12 <2.0 and m12≥2.0

| m12 | 0÷2.0 | ≥2.0 | All |
|---|---|---|---|
| E,L (T=-7÷-1) | 269(264.9) 0.975 | 7(11.1) 0.025 | 276 |
| S,I (T=0÷20) | 779(783.1) 0.955 | 37(32.9) 0.045 | 816 |
| All | 1048 0.960 | 44 0.040 | 1092 |

Table 4 gives the corresponding data for second ranked galaxies for the cases m12 <2.0 and m12≥2.0. From the table it can be seen that in the groups with the magnitude gapes m12≥2.0 among the second ranked galaxies there is not observed large relative number of elliptical and lenticular galaxies in comparison with the expected one also. It seems that the opposite phenomenon is observed, but its statistical significance is small (α≈0.2).

Thus, it can be confidently asserted that the relative number of elliptical and lenticular galaxies among the first and second ranked galaxies does not increase with respect to the expected when the gap m12 increases (when it is assumed that there is no connection between the morphological types the first and second ranked galaxies and the magnitude gap m12). This contradicts the version of the merger. As concern то the hypothesis proposed by Ambartsumian about the origin of galaxies due to the division of super-dense proto-stellar matter, there is no contradiction here. However, more research is needed. The followings are possible topics for farther investigation: detailed morphological study of the first and second ranked galaxies, the study of dynamical conditions in groups, the study of X-ray, infrared, radio emission of groups and these galaxies, star formation, etc. We will carry out this research in the future.

*4. Conclusion.* In the present study, the dependences of the morphological types of the first and second ranked group galaxies on the magnitude gap were studied. This value can characterize the dynamical age of the group. If we accept the hypothesis of galactic mergers, a large gap will characterize a large dynamical age, that is, we are dealing with a relatively old system. If we accept the hypothesis of the origin of galaxies from a protogroup matter by disintegration, proposed by Ambartsumian, then a large gap either has nothing to do with dynamical age or corresponds to a lower dynamic age, that is, perhaps we are dealing with a younger system.

Galactic merging leads to a morphological change in the brightest galaxy of the group, it turns into an elliptical galaxy. The Ambartsumian's hypothesis does not imply a morphological change the brightest galaxy of the group.

The above statistical study shows that there is no increase in the relative number of elliptical galaxies among the first and second ranked group galaxies with a large magnitude gaps (in comparison with the expected, assuming that the morphological type of these galaxies does not depend on the magnitude gap). This result contradicts the merger hypothesis. The hypothesis proposed by Ambartsumian does not contradict this result.